# Un Sistema de Escritura de Traductores de Escritura Vía Web

**José Daniel Texier**
Universidad Nacional Experimental del Tachira
San Cristobal, Venezuela
dantexier@gmail.com
jtexier@unet.edu.ve

**Manuel Bermúdez**
University of Florida
Gainesville, Estados Unidos
manuel@cise.ufl.edu

### RESUMEN

Un compilador es un programa desarrollado en un lenguaje de programación que lee un archivo llamado programa fuente. Luego lo traduce y lo convierte en otro programa llamado objeto, o bien en su defecto genera una salida. Por tal motivo, una manera correcta de entender cualquier lenguaje de programación es analizando su proceso de compilación, el cual es muy similar entre todos los paradigmás o enfoques de programación existentes. Se desea generar una herramienta que permita el aprendizaje en un curso universitario, de todo el proceso de compilación que ocurre en cualquier lenguaje de programación, sin importar la plataforma donde se aplique, logrando una mejor comprensión y una aplicación más inmediata. Para ello se ha desarrollado una aplicación Web, la cual al unirla con el compilador conforman un Sistema de Escritura de Traductores. Este sistema completo, permite extender y modificar el compilador. El sistema propuesto se implemento a través de un modulo en Moodle que permite interactuar con el compilador escogido por el coordinador del curso. Moodle es un sistema de gestión de curso libre (Course Management System, CMS) que ayuda a los educadores a crear comunidades de aprendizaje en línea. El software será distribuido bajo la licencia de software libre GPL.

**Palabras claves:** compilador, Moodle, educación a distancia, software libre.

### ABSTRACT

A compilator is a program which is development in a programming language that read a file known as source. After this file have to translate and have to convert in other program known as object or to generate a exit. The best way for to know any programming language is analizing a compilation process which is same in all programming paradigm existents. To like to generate a tool that permit a learning in university course. This course could explain in any plataform such as Linux o Windows. This goal is posible through development a Web aplication which is unite with a compilator, it is Traductor Writing System (Sistema de Escritura de Traductores). This system is complete and permit extend and modify the compilator. The system is a module in Moodle which is a Course Management System (CMS) that help teachers for to create comunities of learning in line. This software is in free software license (GPL).

**Keywords:** compilator, Moodle, continuing education, free software.

## 1. INTRODUCCION

Actualmente en el mundo de la informática cuenta con aproximadamente 300 sistemás operativos y más de 1500 lenguajes de programación, aunque los más usados son aproximadamente 6 sistemás operativos y 14 lenguajes de programación (MCT, 2004).

Estos lenguajes en general son implementados por un compilador. Un compilador es un programa desarrollado en un lenguaje de programación que lee un archivo llamado programa fuente. Luego lo traduce y lo convierte en otro programa llamado objeto, o bien en su defecto genera una salida (Aho, 1990). Una manera apropiada de entender cualquier lenguaje de programación es analizando su proceso de compilacion.

Actualmente existen cuatro paradigmás en el mundo de los lenguajes de programación: imperativo, orientado a objetos, lógico y funcional (Bermúdez, 2006). Cada paradigma de programación tiene muy bien definida su metodología de trabajo, a fin de lograr sus objetivos especificos (Bermúdez, 2006). El paradigma imperativo es el más utilizado y mejor desarrollado actualmente (Bermúdez, 2003a y 2003b. Tucker, 2003). Emergio junto con las primeras computadoras, sus programás en los años cuarenta, y sus elementos reflejan directamente las características de la arquitectura de las computadoras modernas. El paradigma orientado a objetos (Tucker, 2003) es un estilo de programación que involucra metodologías y técnicas para el modelado y desarrollo de software basado en la construcción y conexión de objetos. Una visión orientada a objetos incluye entre otros aspectos el analisis, diseño y codificación, basado en cuatro principios fundamentales: la abstraccion de datos, el encapsulamiento de los datos en clases, el ocultamiento de información, y el polimorfismo de los objetos. El paradigma lógico (Tucker, 2003) (tambien es conocido como programación declarativa), apareció como un paradigma independiente en los años setenta. Se distingue de otros paradigmas porque obliga al programador declar los objetivos o hechos, en lugar de dar un algoritmo con el que se consiguen dichos objetivos. Los hechos se expresan como un grupo de reglas para obtener lo propuesto. Por esta razon a este paradigma se le conoce como la programación basada en reglas. Finalmente, el paradigma funcional (Tucker, 2003) apareció a principios de los años sesenta. Su característica esencial es que los cálculos se ven como una función matemática. A diferencia de la programación imperativa, no hay un concepto explcito del estado de la memoria y por tanto no hay necesidad de una instruccion de asignación.

El presente trabajo tuvo como objetivo, desarrollar e implementar un sistema que facilite al usuario, el aprendizaje del proceso de compilacion de un lenguaje de programación. El mismo se realizó vía Web, donde el coordinador del curso puede desde cualquier parte del mundo supervisar el trabajo de cada uno de los usuarios del mismo. Además, esta alternativa representa una ventaja hoy en día, ya que las universidades nacionales Venezolanas deben tener independencia de la plataforma (Bermúdez, 2003).

### PLANTEAMIENTO DEL PROBLEMA

Con este proyecto, se deseo generar una herramienta que facilite el proceso de aprendizaje del proceso de compilación, sin importar la plataforma en la cual se aplique. Para ello se planteo desarrollar una aplicación Web que permita extender y modificar un compilador de una forma interactiva y totalmente independiente de la plataforma que se use. Esta aplicación mostró cada uno de los subprocesos de compilación en forma secuencial, donde cada cambio que se realice debe influir inmediatamente en los subprocesos siguientes y el usuario puede percibirlos, permitiéndole un mejor aprendizaje.

A continuación se exponen de forma resumida los subprocesos que conforman el proceso de compilación:

- Análisis léxico: el proceso de analizar caracteres del programa fuente, formando unidades léxicas llamadas token.
- Análisis sintáctico: en este subproceso se identifica las estructuras sintácticas de la secuencia de tokens.
- Análisis de semántica estática: una vez establecida la estructura sintáctica producida por el analizador léxico del programa fuente, se analiza el significado de ese código, incluyendo análisis de compatibilidad de tipos de datos, declaración de variables, etc.
- Generación de código intermedio: en este subproceso se genera una versión intermedia de código semánticamente equivalente al código fuente.
- Optimización de código: en esta etapa se intenta mejorar el código intermedio, con el objetivo de producir un código de máquina más rápido de ejecutar, o bien con menor tamaño.

- Generación de código: en esta etapa se genera el código objeto, que por lo general consiste de código de maquina relocalizable o lenguaje ensamblador.

Bermúdez sugiere que a partir del trabajo realizado por él (Bermúdez, 2006), se desarrolle lo propuesto en esta tesis en un entorno Web soportado por herramientas de última generación, para poder mejorar muchos aspectos que limitan la versión del año 2003 (Bermúdez, 2006), tales como:

- El control de la aplicación no depende totalmente del instructor.
- No permite el acceso a varios usuarios, es decir, la aplicación es monousuario.
- La aplicación se debe instalar en el sitio donde cada uno realice su sesión práctica.
- No es interactivo ni atractivo para el usuario.
- No se presta para dar soporte a cursos a distancia.
- El nuevo sistema amplía y fortalece los cursos o seminarios de compiladores e interpretes y también tiene la ventaja de usar esta herramienta para educar a distancia.

## 2. METODOLOGÍA

El desarrollo del trabajo se efectuó de acuerdo al modelo de Ingeniería Web (IWeb) propuesto por Pressman (Pressman, 2002), donde indica que la IWeb demanda un proceso de software incremental y evolutivo. Pressman también señala que el modelo en las primeras versiones puede ser un modelo en papel o un prototipo, y durante las últimas iteraciones se producen versiones cada vez más completas del sistema diseñado. La IWeb se divide en un número de actividades estructurales, también llamadas regiones de tareas (Figura 1). Generalmente, existen entre tres y seis regiones de tareas, las cuales no necesariamente se deben aplicar todas por cada iteración.

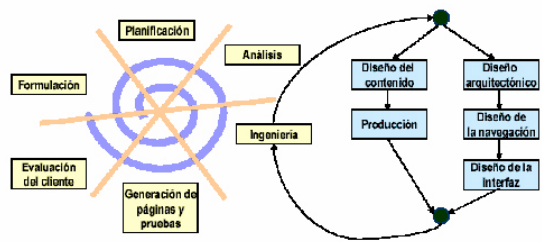

Figura 1. Modelo del Proceso IWeb.

El proceso IWeb comienza con la formulación, que es la actividad que identifica las metas y los objetivos de la aplicación Web (WebApp) a desarrollar, y establece el ámbito del incremento, es decir, incremento del software por cada etapa.

La planificación estima el costo global del proyecto, evalúa los riesgos asociados con el esfuerzo del desarrollo y define una planificación del desarrollo bien granulada para el incremento final de la WebApp.

El análisis establece los requisitos técnicos para la WebApp e identifica los elementos del contenido que se van a incorporar. También se definen los requisitos del diseño gráfico.

La actividad de ingeniería incorpora dos diseños principales. El primero se refiere al diseño del contenido y la producción, y que de ser necesario, será desarrollado por personas no pertenecientes al desarrollo del sistema, es decir, diseñadores gráficos. Se lleva a cabo la integración de las tareas referentes al diseño arquitectónico (estructura global de hipermedia), diseño de navegación (permiten al usuario acceder al contenido y a los servicios) y diseño de la interfaz (impactar con la primera impresión).

La generación de páginas es una actividad de construcción que hace mucho uso de las herramientas automatizadas para la creación de la WebApp. El contenido definido en la actividad de ingeniería se fusiona con los diseños arquitectónicos, de navegación y de la interfaz para elaborar páginas Web ejecutables en HTML, XML y otros lenguajes orientados al proceso. Durante esta actividad también se lleva a cabo la integración con el software intermedio, conocido como middleware. Las pruebas ejercitan la navegación, intentan describir los errores de applets, guiones y formularios, y ayuda a asegurar que la WebApp funcionara correctamente en diferentes entornos (navegadores).

Cada incremento producido como parte del proceso IWeb se revisa durante la actividad de evaluación del cliente. Es en este punto en donde se solicitan cambios. Estos cambios se integran en la siguiente ruta mediante el flujo incremental del proceso.

### IMPLEMENTACIÓN DEL MODELO

En esta investigación se realizaron cuatro iteraciones. En cada una se deben aplicar las seis actividades, pero las mismas estarán sujetas al ámbito definido en la formulación del problema. Se describieron las diferentes actividades del modelo IWeb en consonancia a los objetivos definidos. Las cuatro (4) iteraciones fueron las siguientes:

- Primera iteración, tuvo como resultado un trabajo en papel.

- Segunda iteración, tuvo como resultado un prototipo del sistema.

- Tercera iteración, se realizo un sistema con todas los subprocesos de un compilador.

- Cuarta iteración, se incluyo el sistema realizado en la iteración anterior en Moodle, obteniendo una aplicación robusta.

Las actividades de acuerdo al modelo IWeb, que permitieron desarrollar el sistema que esta siendo usado en el curso de Compiladores en la Universidad Nacional Experimental del Táchira y será usado en la Universidad de Florida, son:

- Formulación: se identificaron las metas y objetivos del sistema para establecer el ámbito de las cuatro iteraciones aplicadas.

- La planificación: se evaluaron los riesgos asociados con el esfuerzo del desarrollo, y el tiempo de ejecución de la misma.

- El análisis: permitió establecer los requisitos técnicos del sistema y se identificaron los elementos del contenido tomando como base el software de consola desarrollado anteriormente. También se tomaron en cuenta los requerimientos del diseño gráfico.

- Ingeniería: se logró la integración del diseño arquitectónico, de navegación y de interfaz.

- Generación de páginas y pruebas: se fusiono el análisis y diseño del sistema con la escogencia del lenguaje de programación Web PHP. Además se realizaron pruebas basadas en la metodología IWeb.

- Evaluación del cliente: permitió corregir errores por cada iteración, obteniendo como resultado una evolución en comparación con la iteración anterior.

- En el próximo capítulo se describirán al detalle las diferentes actividades vinculadas a cada una de las iteraciones de acuerdo a la metodología IWeb.

## 3. DESCRIPCIÓN DE LA APLICACIÓN WEB

La aplicación desarrollada esta definida en diferentes fases. Estas fases y subfases se clasificaron de la siguiente manera:

- Fase 1 - Compiler. Se especifica el compilador, en este modulo se definen las reglas de juego, es decir, como funcionara el analizador léxico (a través del comando LEX en GNU/Linux) y el analizador sintáctico (a través del comando YACC en GNU/Linux). También permite definir como será el proceso de análisis de semántica estática y la generación de código.
- Fase 2 - Run. Se corre el compilador, a partir de un programa fuente, se obtienen sus tokens, un árbol sintáctico, su semántica y el código de maquina.
- Fase 3 - Interpreter. Se corre el programa compilado, este modulo permite ejecutar el código de máquina generado y mostrar la secuencia de pasos de la ejecución del compilador.

Distribución general de la aplicación Web:
- Fase 1: Compiler.
    - Scanner.
    - Parser.
    - Contrainer.
    - Generator.
- Fase 2: Run.
    - Source.
    - Scanning.
    - Parsing.
    - Constrain.
    - GenCode.
- Fase 3: Interpreter.
    - Code.

En el gráfico que se presenta a continuación (figura 2), se observa el diseño del compilador donde se destacan las tres fases principales del diseño de la aplicación Web desarrollada. Pero antes se destaca, que el sistema cuenta con los subproceso llamados: *flex, gcc, pgen y yacc*. El *flex y yacc*, son herramientas que permiten generar salidas léxicas y sintácticas de acuerdo a las reglas definidas para estos procesos (Stallamn, 2004). El *pgen* es un subproceso definido por Bermúdez (Bermúdez, 2003a) que permite adaptar la salida de las reglas sintácticas al *yacc*. Y el *gcc* es el compilador de C para GNU/Linux (Stallamn, 2004).

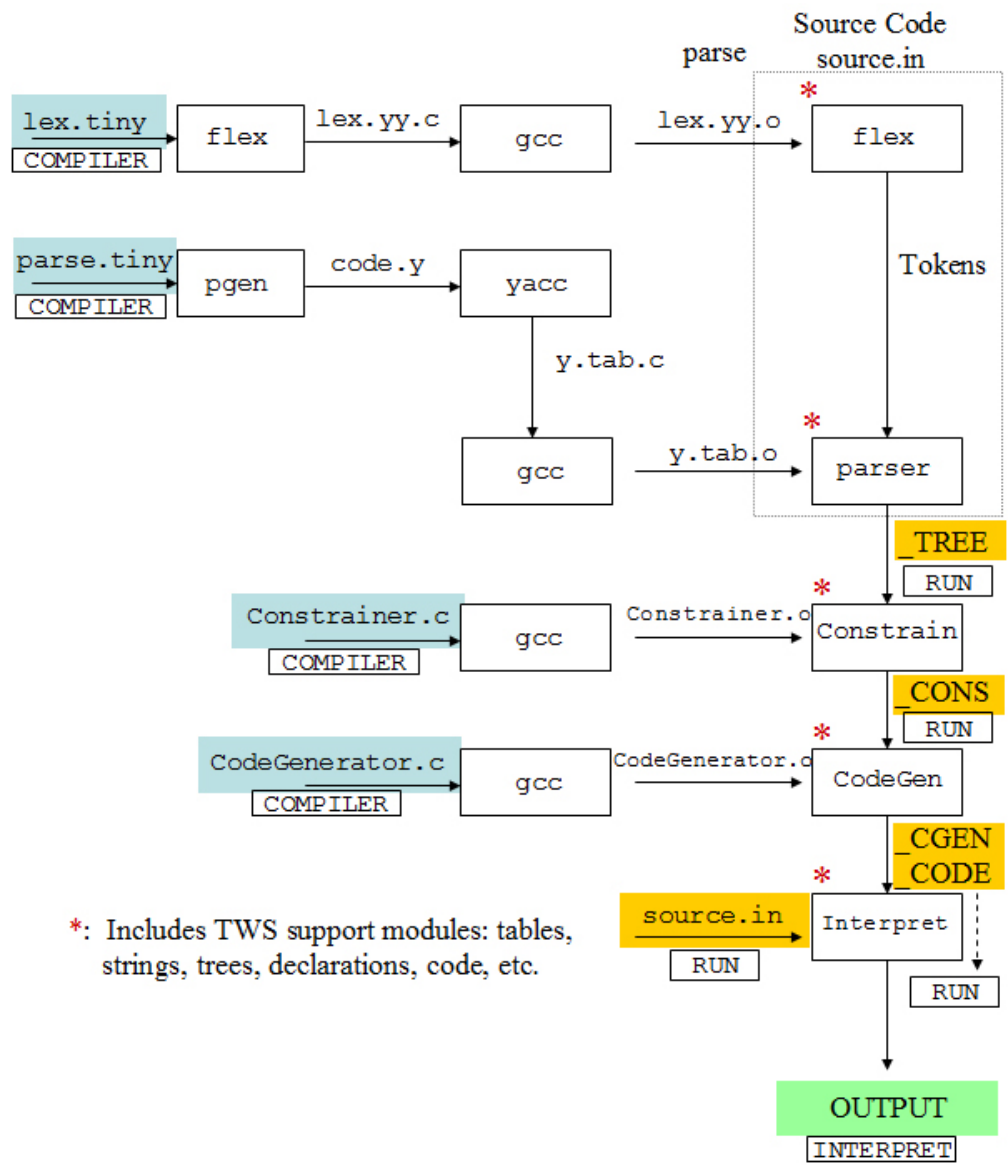

Figura 2. Estructura Interna del Compilador – Modo Consola.

La primera fase es conocida como COMPILER, se encuentra de color azul claro (cada una representa una subfase de COMPILER), la segunda fase llamada RUN esta de color amarillo (son cinco nombres de archivos que representan cada subfase de RUN) y la ultima fase que tiene como nombre INTERPRET es de color verde claro (esa es la única subfase de INTERPRET). También se observa que los rectángulos representan entidades de software, es decir, programa que se van a ejecutar tales como: *lex o flex, yacc, gcc*, entre otros. Pero también hay programa objetos generados por el compilador y que llevan como nombre: pgen, parser, Constrain, CodeGen, Interpretet.

Después de entender el funcionamiento interno de la aplicación Web, se explicara los componentes de la misma basadas en un diseño acorde a cumplir los objetivos planteados para la enseñanza de un curso de Compiladores e Interpretes (Bermúdez, 2003a). Las tres modalidades son:

- Modalidad de edición del compilador (Compiler).
- Modalidad de compilación del compilador (Run).
- Modalidad de ejecución del compilador (Interpret).

Finalmente se presenta el sistema desarrolla en la siguiente figura:

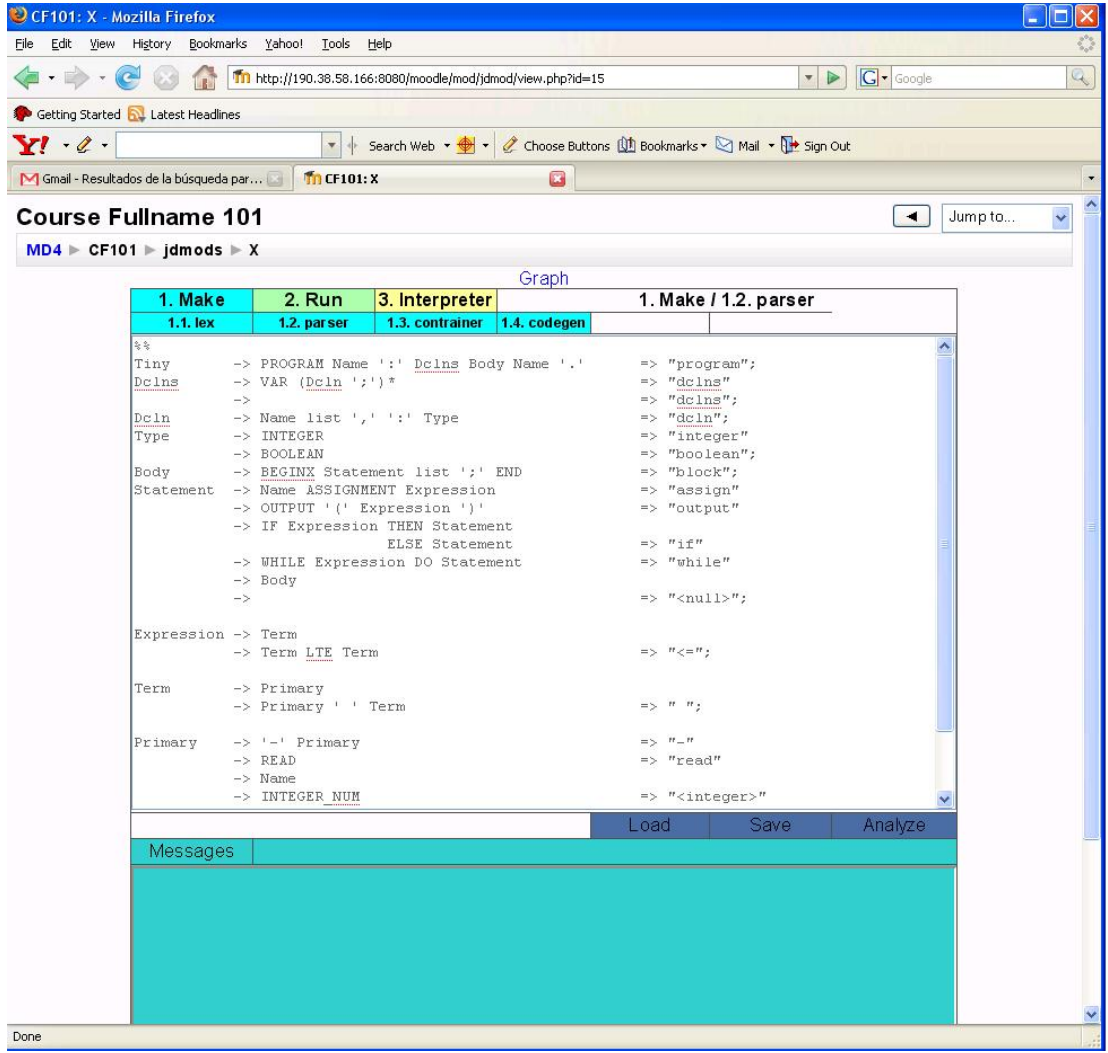

Figura 3. Sistema de Traductores de Escritura.

## 4. CONCLUSIONES

- El Sistema de Escritura de Traductores desarrollado se caracteriza por estar implementado bajo un ambiente Web, donde no importa la plataforma donde se encuentre en cliente (navegador). Siempre que se tenga acceso desde Internet (servidor con acceso a Internet), el usuario podrá hacer uso de la herramienta desde cualquier parte del mundo. El modelo del compilador desarrollado se implemente bajo el paradigma de programación imperativo, que hace referencia a un lenguaje de recorrido completo que soporta las siguientes características: tipos de datos, estructuras de control, asignación de elementos, comandos de entradas y salidas, y procesamiento de caracteres, entre otras.

- Una vez analizado el Sistema de Escritura de Traductores desarrollado por Bermúdez, se implemento un sistema va Web bajo un modelo lógico adaptado al sistema original. Este modelo lógico contiene en tres fases. La primera permite al usuario implementar las reglas léxicas, sintácticas, semánticas y la generación de código. La segunda fase permite compilar el programa fuente deseado y generar los archivos correspondientes a las subfases anteriores. La última fase permite ejecutar el código objeto generado.

- Ante el deseo de contribuir al desarrollo de Software Libre en el mundo y acogiéndonos a los lineamientos del gobierno Venezolano, el desarrollo del proyecto se realizo bajo la filosofía del Software Libre, logrando obtener un producto académico de calidad y fortaleciendo la lucha en las Universidades Venezolanas por promover el Software Libre.

- El proyecto realizado se ejecuto en cuatros fases. En cada una de ellas se realizaron las pruebas pertinentes de acuerdo a la metodología aplicada, alcanzando el objetivo propuesto: una herramienta Web que mejore la pedagoga en la enseñanza de los Compiladores. Siguiendo la filosofía del Software Libre, el acceso al código fuente esta disponible en un archivo ubicado en la sección de archivos del curso creado en Moodle, as como en la página personal del Dr. Bermúdez y del autor de este proyecto.

## 5. Autorización y Renuncia

Las siguientes palabras aparecerán en la sección Autorización y Renuncia al final del documento: "Los autores autorizan a LACCEI para publicar el escrito en los procedimientos de la conferencia. LACCEI o los editores no son responsables ni por el contenido ni por las implicaciones de lo que esta expresado en el escrito."

## 6. Entrega

Envíe todos los manuscritos electrónicamente a través de la página para envío de escritos de la conferencia en el link que aparece en la página Web de LACCEI: http://www.laccei.org en formato Microsoft Word. Todos los manuscritos deben ser recibidos por LACCEI antes de la fecha limite de Abril 15, 2007 que ha sido establecida para recibir los manuscritos.

## Referencias

ACM. Computing Curricula, Final Draft. 15 Diciembre del 2001. http://www.computer.org/education/cc2001/~nal.
Alfred Aho, Ravi Sethi y Jefrey Ullman. Compiladores - Principios, técnicas y herramientas. Primera edición. Editorial Addison Wesley. Wilminton, Estados Unidos. 1990.
Manuel Bermudez. Course Principles of Programming Languages. Pagina Web, Septiembre 2006. http://www.cise.uf.edu/class/cop5555su06/index.html
Manuel Bermúdez. Modernización de la enseñanza de la Traducción, XI Congreso Iberoamericano de Educación Superior en Computación CIESC-03. September 29 - October 3, La Paz, Bolivia.
Manuel Bermúdez. Traductor Writing System (TWS). Pagina Web. 2003. http://www.cise.uf.edu/class/cop5555su06/language.notes.html
Ministerio de Ciencia y Tecnóloga (MCT). Libro Amarillo del Software Libre. Primera edición. MCT. Caracas, Venezuela. 2004.
Moodle - Sistema para la Administración de Cursos. http://www.moodle.org. 2007
Richard Stallman. Software libre para una sociedad libre. Traficantes de Sueños. España. 2004.
Roger Pressman. Ingeniera del Software. McGraw Hill. Quinta edición. Madrid, España. 2002.
Tucker A. y Noonan R. Programming Languages, Principles and Paradigms. Editorial McGraw Hill. Estados Unidos. 2003.
Wikipedia La Enciclopedia Libre. http://es.wikipedia.org/wiki.

*Autorización y Renuncia*

*Los autores authorizan a LACCEI para publicar el escrito en los procedimientos de la conferencia. LACCEI o los editors no son responsables ni por el contenido ni por las implicaciones de lo que esta expresado en el escrito*

*Authorization and Disclaimer*